\documentclass[prb,twocolumn,showpacs,preprintnumbers,amsmath,amssymb]{revtex4}

\usepackage{epsfig}
\usepackage{dcolumn}
\usepackage{bm}

\begin{document}

\preprint{}

\title{\bf Mobility of thorium ions in liquid xenon}

\author{K.~Wamba,$^1$ C.~Hall,$^1$ M.~Breidenbach,$^1$ R.~Conley,$^1$ A.~Odian,$^1$ C.Y.~Prescott,$^1$ P.C.~Rowson,$^1$ 
J.~Sevilla,$^1$  K.~Skarpaas,$^1$ R.~DeVoe,$^2$ Z.~Djurcic,$^4$\protect\cite{byline3} W.M.~Fairbank Jr.,$^5$ 
G.~Gratta,$^2$ M.~Green,$^2$ K.~Hall,$^6$ M.~Hauger,$^6$ S-C.~Jeng,$^5$ T.~Koffas,$^2$\protect\cite{byline4} 
F. LePort,$^2$ D.~Leonard,$^4$ J-M.~Martin,$^6$ R.~Neilson,$^2$ L.~Ounalli,$^6$  
A.~Piepke,$^4$ D.~Schenker,$^6$ V.~Stekhanov,$^3$ J-L.~Vuilleumier,$^6$ S.~Waldman,$^2$ P.~Weber,$^6$ J.~Wodin$^2$}
	 
\newcommand{\commuf}{$,$}

\affiliation{ $^1$Stanford Linear Accelerator Center\commuf ~Menlo Park CA 94025\commuf ~USA\\
$^2$ Physics Department\commuf ~ Stanford University\commuf ~ Stanford CA 94305\commuf ~ USA\\
$^3$ Institute for Theoretical and Experimental Physics\commuf ~ Moscow\commuf ~ Russia\\ 
$^4$ Department of Physics and Astronomy\commuf ~ University of Alabama\commuf ~ Tuscaloosa AL 35487\commuf ~ USA\\ 
$^5$ Physics Department\commuf ~ Colorado State University\commuf ~ Fort Collins CO 80523\commuf ~ USA\\
$^6$ Institut de Physique\commuf ~ Universite de Neuchatel\commuf ~ Neuchatel\commuf ~ Switzerland\\  
}
\collaboration{EXO Collaboration}
\date{\today}

\date{\today}

\begin{abstract}
We present a measurement of the 
$^{226}$Th ion mobility in LXe at 163.0 K and 0.9 bar.  The result 
obtained, 0.240$\pm$0.011 (stat) $\pm$0.011 (syst) cm$^{2}$/(kV-s), is compared with a popular model of ion transport.   
\end{abstract}

\maketitle

\section{\label{sec:level1}Introduction}
Free ions in the liquefied noble gases at low concentrations represent
a theoretically straightforward condensed matter system.  This is largely because of the liquid's 
atomic nature, implying relatively few microscopic internal 
degrees of freedom.\cite{dav}
Despite this fact, the experimental study of these systems has 
long been confounded by the difficulty in reproducibly generating and transporting ions of known identity while 
controlling for the effects of impurity ions.  Accordingly, there have been relatively few
transport studies for ions in the liquefied noble gases other than LHe
reported in the literature to date.  Furthermore, studies that deal only with molecular ions,\cite{dav,sch,pal,dav2,hfk}  
are more common  than those that deal with monatomic species.\cite{wal,fai}  
There is therefore a clear need for additional data on monatomic ion transport particularly in the heavy noble gases.  
In the present work we consider the system of monatomic Th ions in LXe.\\ \indent
This research also has important applications to the physics of ionizing particle detectors, 
especially in the area of rare event searches that use a liquefied noble gas as the detection medium.\cite{moe,geo,suz} 
For these systems, the transport properties of Th and
U ions are of concern because these elements in trace amounts represent a large source of background radioactivity.  It is 
therefore desirable to understand how electric fields might be used to draw these ions out of the detection medium.\cite{suz}  
For LXe-based detectors, Cs and Ba ions are also interesting because their respective separation from the detection medium 
and subsequent identification can greatly enhance the sensitivity to solar neutrinos and to double-beta 
decay.\cite{moe,geo}  The motivation for the present study is work involving Ba ion retrieval from LXe as part of R\&D for a 
large double-beta decay experiment.\cite{dan}\\ \indent
In this paper, we report on the measurement of the mobility of $^{226}$Th ions in liquid Xe at 163.0 K and 0.9 bar pressure.  
We also compare our results and those of other workers with the theoretical calculations of Hilt, Schmidt and Khrapak.\cite{sch}\\ 
\indent
\section{Experiment}

\begin{figure} 
\includegraphics[width=8cm]{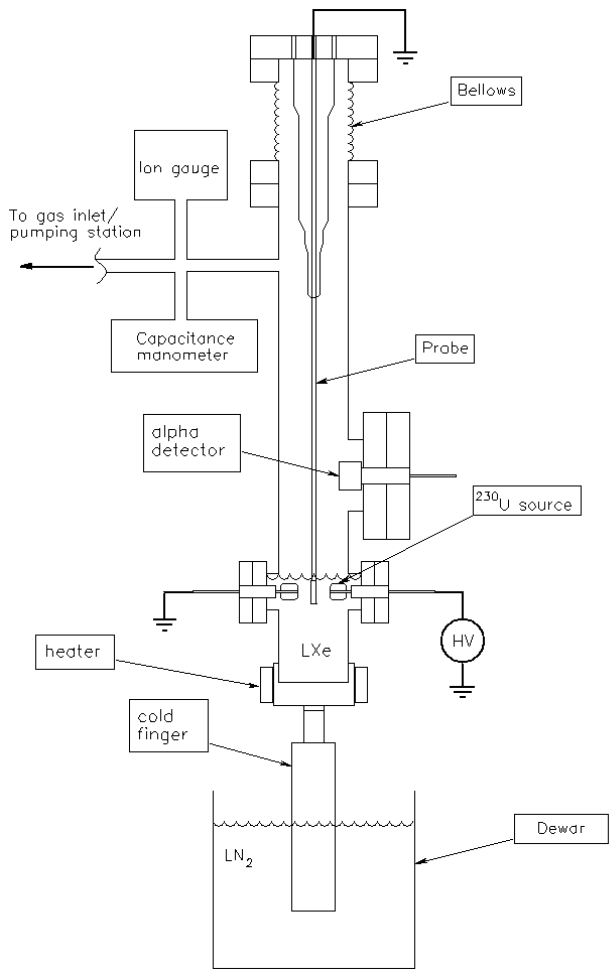} 
\caption{\label{fig:system}
 Schematic of the probe cell.}
\end{figure}

Our system, schematically shown in Fig. \ref{fig:system}, consists of a vacuum-insulated cell which we fill with about 10 cm$^{3}$
 of LXe.The cell is cooled by a cold finger that forms a thermal link with a liquid nitrogen reservoir.  The temperature 
is maintained at 163.0$\pm$0.1 K 
by a resistive heater.  The cell, built almost entirely out of UHV-compatible materials,
is baked at 400 K for several days, after which it is pumped to a ~7.0$\times$10$^{-8}$ mbar vacuum by means of two 
turbomolecular pumps.  When filling the cell, we pass research-grade Xe (which, as supplied, is quoted as 99.999\% pure) through 
a hot Zr getter,\cite{sae} which is rated to remove reactive impurities to the $<$1 ppb level.  These precautions are taken 
to minimize any effect of impurities that might influence ion transport characteristics.\\ \indent
Th ions are produced by a $^{230}$U source mounted inside the cell.  The 
\begin{math}^{230}\end{math}U was obtained as a decay product of \begin{math}^{230}\end{math}Pa, in turn produced in the reaction 
$^{232}_{90}$Th+p$\rightarrow$$^{230}_{91}$Pa+3n on
 natural thorium at a cyclotron.\cite{tew}  After irradiation, the \begin{math}^{230}\end{math}Pa 
was left to $\beta$-decay for a few of its 17.4 day half lives, allowing some build-up of \begin{math}^{230}\end{math}U, which was 
then
chemically separated and electroplated onto a platinum disk. Fig. \ref{fig:levelscheme} shows the decay scheme for the final 
$^{230}$U source.\\ \indent
\begin{math}^{226}\end{math}Th ions are formed as a result of the 20.8 day \begin{math}\alpha\end{math}
 decay of \begin{math}^{230}\end{math}U.  The average recoil
energy of $\sim$100 keV extracts $^{226}$Th atoms from the surface in a highly ionized state.  As the   
\begin{math}^{226}\end{math}Th ion enters the liquid it exchanges charge with the surrounding Xe atoms and reaches equilibrium 
in some final charge state.  Although the energetics of the charge exchange in principle favor a final
 state containing Xe\begin{math}^{+}\end{math} and Th\begin{math}^{+}\end{math} (Ref. \onlinecite{wal}), a large number of the
 emerging Th ions can also recombine with free ionization electrons that arise from the original \begin{math}\alpha\end{math}
decay.\cite{wal}  This effect presents a major problem as it greatly suppresses the ion yield.\\ \indent
We collect ions on a stainless steel probe which we lower into the liquid.  To collect ions we place the probe at ground potential
while applying a positive bias to the source electrode.   The tip of the probe is 
fashioned into a 5 mm $\times$ 16 mm, 2 mm thick rectangular stainless steel plate, ensuring
a spatially uniform electric field in the vicinity of the source.  The uniformity of the field is verified by constructing 
a computer model of the electrostatics with the MAXWELL software package.\cite{max}\\ \indent
The ions that land on the probe are detected by retracting the probe vertically to a counting station, located above the liquid 
surface.  It consists 
of a 50 mm\begin{math}^{2}\end{math} active area ion-implanted Si detector\cite{det} that is read out using standard pre- and 
post-amplification electronics.
The ion yield is determined by counting the $\alpha$ particles emitted in the 30.6 min 
\begin{math}^{226}\end{math}Th decay, along with those produced by the subsequent \begin{math}\alpha\end{math} decays of 
\begin{math}^{222}\end{math}Ra, \begin{math}^{218}\end{math}Rn, and 
\begin{math}^{214}\end{math}Po (see Fig. \ref{fig:levelscheme}).\\ \indent
It should be pointed out that in addition to the \begin{math}^{226}\end{math}Th ions, \begin{figure}
\begin{picture}(190,220)(4,-230)
\setlength{\unitlength}{0.5 mm}
\linethickness{0.5 mm}
\put(121,-20){\line(12,0){12}}
\put(121,-20){\vector(-1,-2){10}}
\put(121,-20){\vector(-1,-1){10}}   
\put(75,-25){{\scriptsize 5.82 MeV$\alpha$~30\%}}
\put(115,-37){{\scriptsize 5.89 MeV$\alpha$~70\%}}
\put(122,-26){{\scriptsize $^{230}$U(20.8 d)}}
%
\put(98,-40){\line(12,0){12}} 
\linethickness{0.1mm}
\put(98,-30){\line(12,0){12}}
\linethickness{0.5mm}
\put(98,-40){\vector(-1,-2){10}}
\put(98,-40){\vector(-1,-1){10}}
\put(100,-46){{\scriptsize $^{226}$Th(30.6 m)}}
\put(51,-45){{\scriptsize 6.23 MeV$\alpha$ 23\%}}
\put(92,-56){{\scriptsize 6.34 MeV$\alpha$ 76\%}}
%
\put(75,-60){\line(12,0){12}}
\put(75,-60){\vector(-1,-2){10}}
\linethickness{0.1mm}
\put(75,-50){\line(12,0){12}}
\linethickness{0.5mm}
\put(77,-66){{\scriptsize $^{222}$Ra(38 s)}}
\put(44,-65){{\scriptsize 6.56 MeV$\alpha$}}
%
\put(52,-80){\line(12,0){12}}
\put(52,-80){\vector(-1,-2){10}}
\put(54,-86){{\scriptsize $^{218}$Rn(35 ms)}}
\put(21,-85){{\scriptsize 7.13 MeV$\alpha$}}
%
\put(29,-100){\line(12,0){12}}
\put(29,-100){\vector(-1,-2){10}}
\put(31,-106){{\scriptsize $^{214}$Po(164 $\mu$s)}}
\put(-2,-105){{\scriptsize 7.69 MeV$\alpha$}}
%
\put(6,-120){\line(12,0){12}} 
\put(18,-120){\vector(1,-4){2}} 
\put(20,-124){{\scriptsize $\beta ^-$}}
\put(-15,-126){{\scriptsize $^{210}$Pb(22.3 y)}}
%
\put(17,-129){\line(12,0){12}}
\put(29,-129){\vector(1,-4){2}}
\put(31,-133){{\scriptsize $\beta ^-$}}
\put(-3,-135){{\scriptsize $^{210}$Bi(5.01 d)}}
%
\put(29,-138){\line(12,0){12}}  
\put(29,-138){\vector(-1,-2){10}}        
\put(31,-144){{\scriptsize $^{210}$Po(138.4 d)}}
\put(-4,-147){{\scriptsize 5.30 MeV$\alpha$}}
%
\put(6,-158){\line(12,0){12}}
\put(-10,-164){{\scriptsize $^{206}$Pb(stable)}}
\end{picture}
\caption{\label{fig:levelscheme} Decay scheme for $^{230}$U.  
Decays with branching ratios less than 5\% are not shown.}
\end{figure}
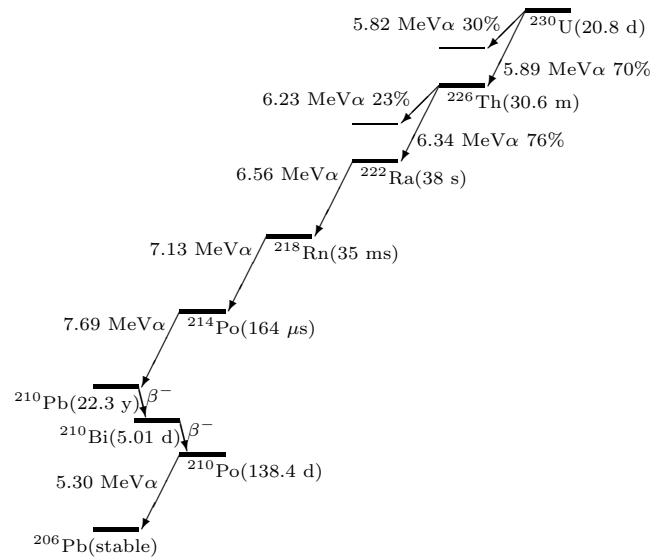  the \begin{math}^{230}\end{math}U source also produces ions arising from the other \begin{math}\alpha\end{math} 
decays in the chain.  These can also drift to the probe and be detected at the counting station.  However, because of their 
considerably shorter lifetimes, they present a negligible background to the \begin{math}^{226}\end{math}Th signal  
provided some short time elapses between the probe extraction and the counting.  In addition, the low angular acceptance of our 
$\alpha$ particle detection system ($\sim$5\%) ensures that the probability of detecting more than one $\alpha$ particle from
the 4-$\alpha$ decay sequence of a given Th atom is negligibly small.\\ \indent
We measure the ion mobility in liquid Xe by determining the ion transit time from source to probe at a given probe distance and 
electric field. This is achieved by biasing the source with an alternating potential produced by a computer driving a high voltage 
amplifier.\cite{tre} The system is set to produce a square wave switching between $V_{drift}$ and $V_{sup}$ with 
respect to the probe, 
\begin{figure}
\includegraphics[width=9cm]{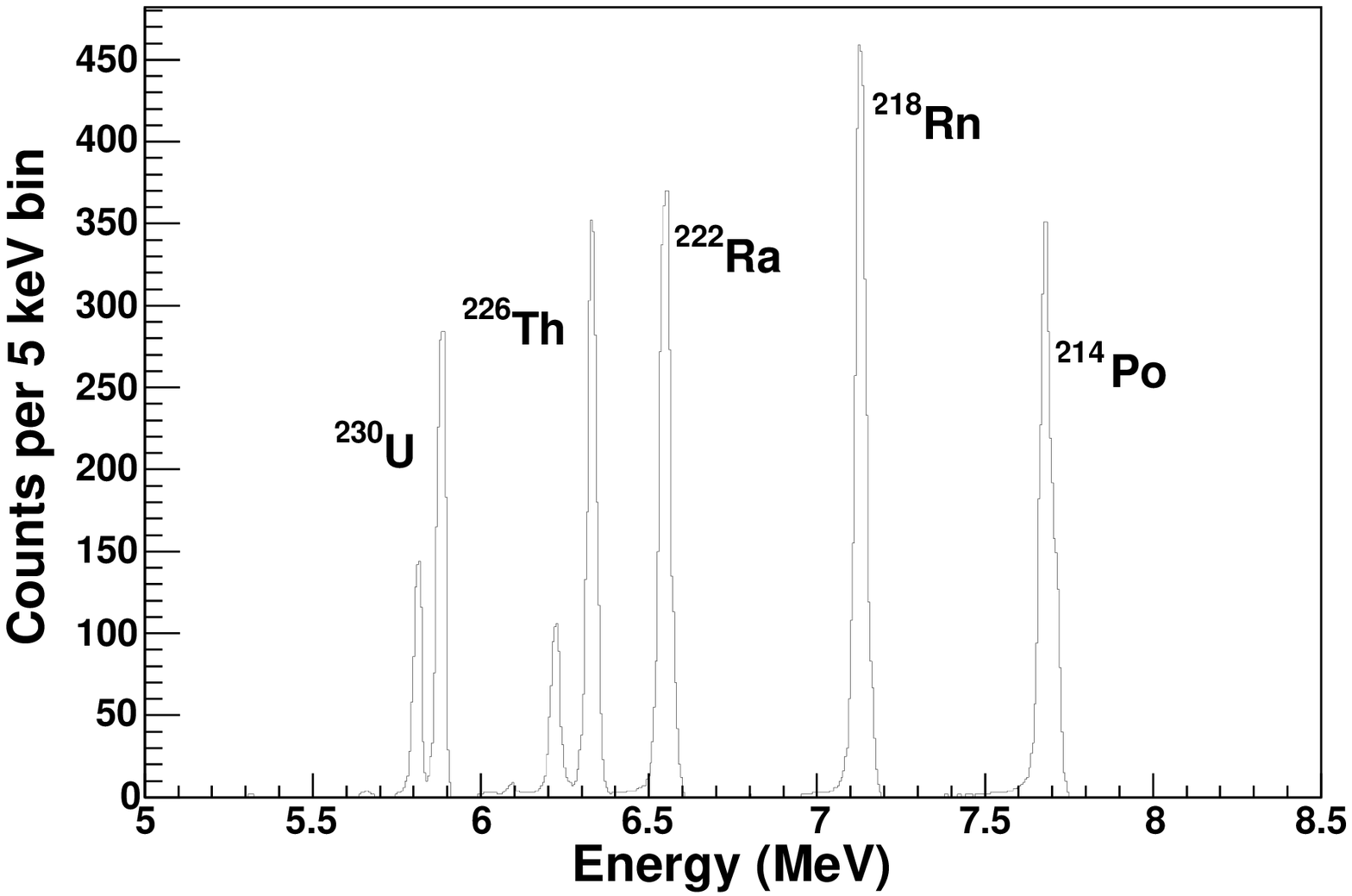}\\
\includegraphics[width=9cm]{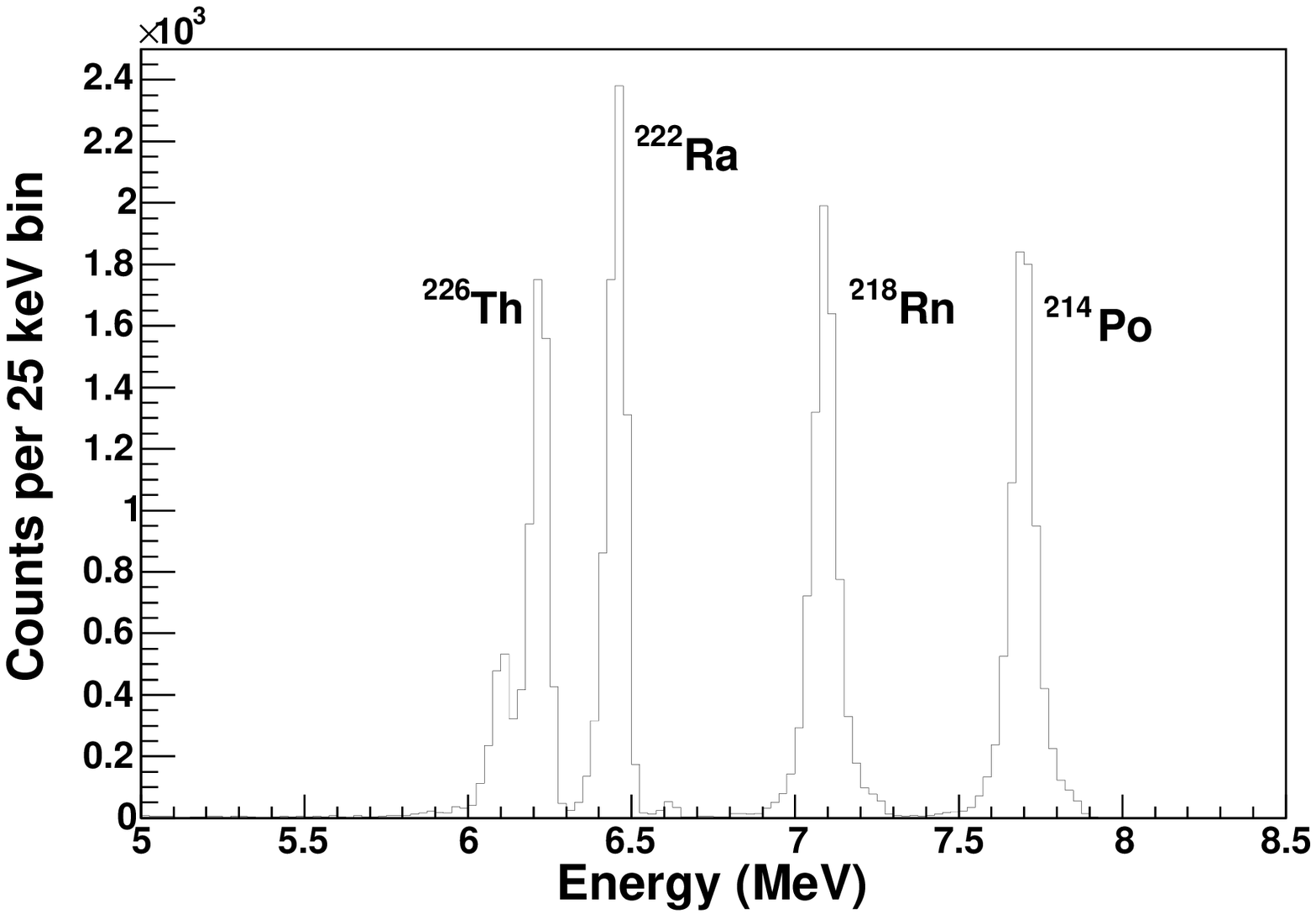}\\
\includegraphics[width=9cm]{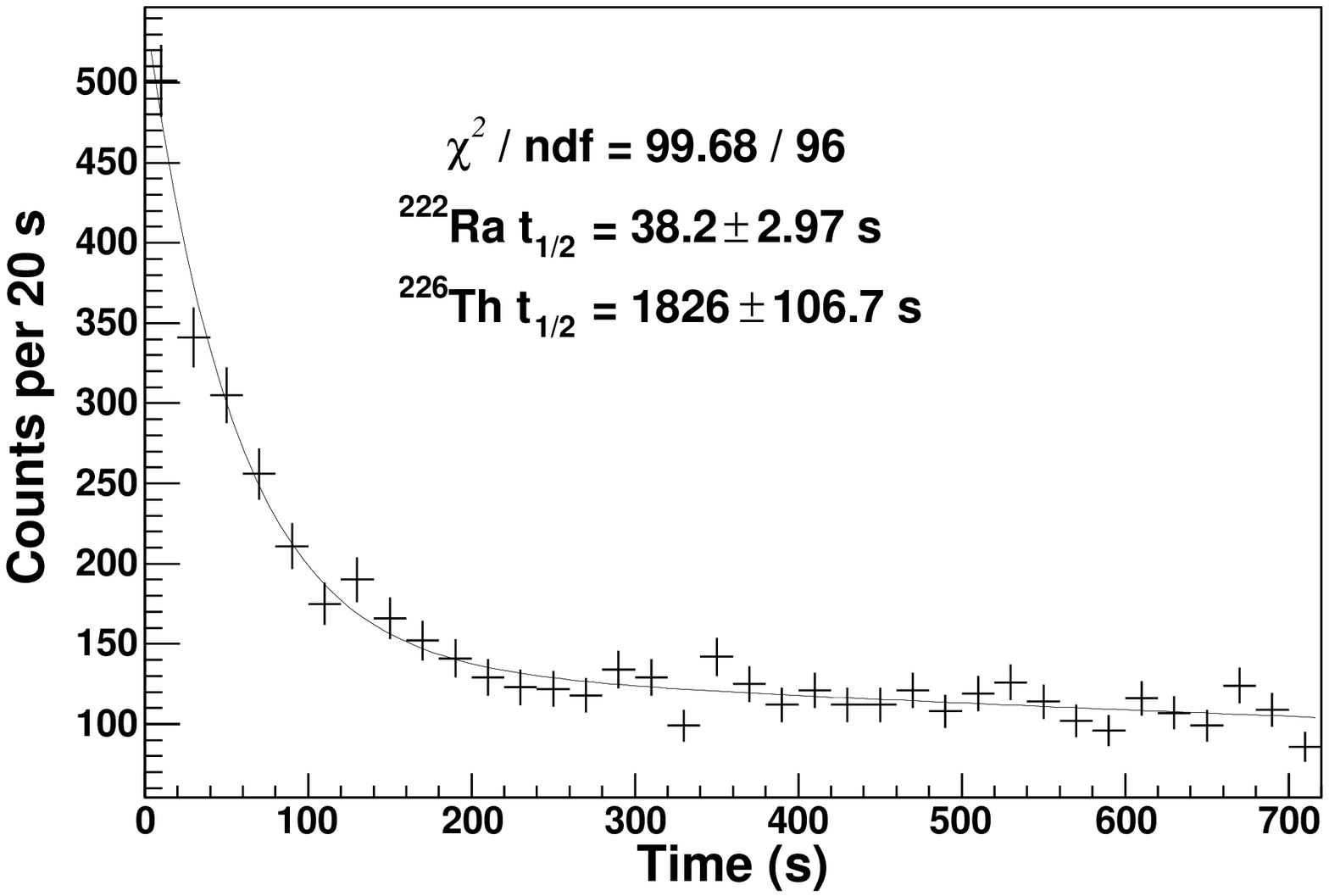}\\
 \caption{\label{fig:alphas} (a) $\alpha$ spectrum of the $^{230}$U source, 
taken under vacuum.\cite{krob}  The $^{230}$U peaks are apparent at 5.82 MeV and 5.89 MeV, while $^{226}$Th is 
identified by the peaks at 6.23 MeV and 6.34 MeV.  $^{222}$Ra, 
$^{218}$Rn, and $^{214}$Po correspond to the peaks at 6.56 MeV, 7.13 MeV, and 7.69 MeV respectively.
(b)  $\alpha$-decay spectrum for ions that are collected on the probe in 
0.2 bar of ambient Xe gas.  Here, the peaks are broadened and shifted to lower energy as 
a result of losses in the Xe gas.  (c) $\alpha$-decay time histogram for the sum over all counts in the alpha
spectrum shown in (b). The best fit is the superposition of two exponentials 
with half-lives of 1826$\pm$106.7 s and 38.2$\pm$2.97 s.  These match the $^{226}$Th and $^{222}$Ra half-lives, whose 
published values are 1830 s and 38 s respectively.\cite{iso}}
\begin{picture}(190,220)(0,0)
\setlength{\unitlength}{0.5mm}
\put(5,361){(c)}
\put(5,482){(b)}
\put(5,604){(a)}
\end{picture}
\end{figure} where $V_{drift}$ can be varied from 1.0 kV to 4.0 kV, and $V_{sup}$ is fixed at -4.0 kV.  $V_{drift}$ is set for a time 
$t_{drift}\approx1$ s, very long compared to the $\sim$20 $\mu$s switching time of the high voltage system.  During each full cycle, the
 voltage is reversed for a time $t_{sup}$.  When this occurs, the ions 
that do not reach the probe during $t_{drift}$ are drawn back towards the source.  This occurs when $t_{drift}$ is too short for the 
ions to complete the journey to the probe.  
We maximize our total ion integration time by having $t_{sup}\le t_{drift}$ and $|V_{drift}|<|V_{sup}|$.\cite{wal}  During each data 
taking run, $V_{drift}$ is set to a particular value and the field is switched for 1800 s.  The constant duration of collection runs 
ensures that the amount of ion loss due to radioactive decay during collection is uniform.
We then measure the ion yield at the counting station over a period of 2 hours to allow sufficient time for the 
collected \begin{math}^{226}\end{math}Th ions to decay before the next run.  Data are taken at four different electric fields, 
using  four source-probe voltages and two source-probe separation distances.\\ \indent
We determine the background by performing 2-hour counts between data runs in which no ions are collected.  In 
these runs the source-probe voltage remains reverse-biased. The observed rate during the background runs 
is averaged over all runs, giving the mean background rate and an associated statistical error.  The resulting
value of 3.16$\pm$0.73 counts per 2-hour run is  
subtracted from the number of counts observed at each $t_{drift}$.  When ions 
are collected, the typical rate is $\sim$30 counts per run.  Sources 
of background include cosmic rays and neutral Th atoms that evaporate from the surface 
of the liquid and diffuse through the gas.\\ \indent
For the purpose of characterizing the system a test in Xe gas was also done. The cell was filled with 0.2 bar of room-temperature Xe 
gas and the probe was alternately brought close to the $^{230}$U source and raised to the counting station, where ions were 
respectively 
collected and 
counted. This was repeated many times over several hours to build up counting statistics.  The electric field 
between source and probe for ion collection was roughly $\sim$200 V/cm for this test.    Shown in  Fig. \ref{fig:alphas} are
the \begin{math}\alpha\end{math} spectrum and time histogram for the collected Th and its decay daughters. 
For comparison, the 
$^{230}$U spectrum, as measured in vacuum at LLNL,\cite{krob} is also shown.  An important conclusion from 
the gas test was that ion recombination with free ionization electrons is a factor of $\sim$200 times more efficient in LXe 
than in gaseous Xe.\\ \indent

\section{Results}
\begin{figure*}
\includegraphics[width=18cm]{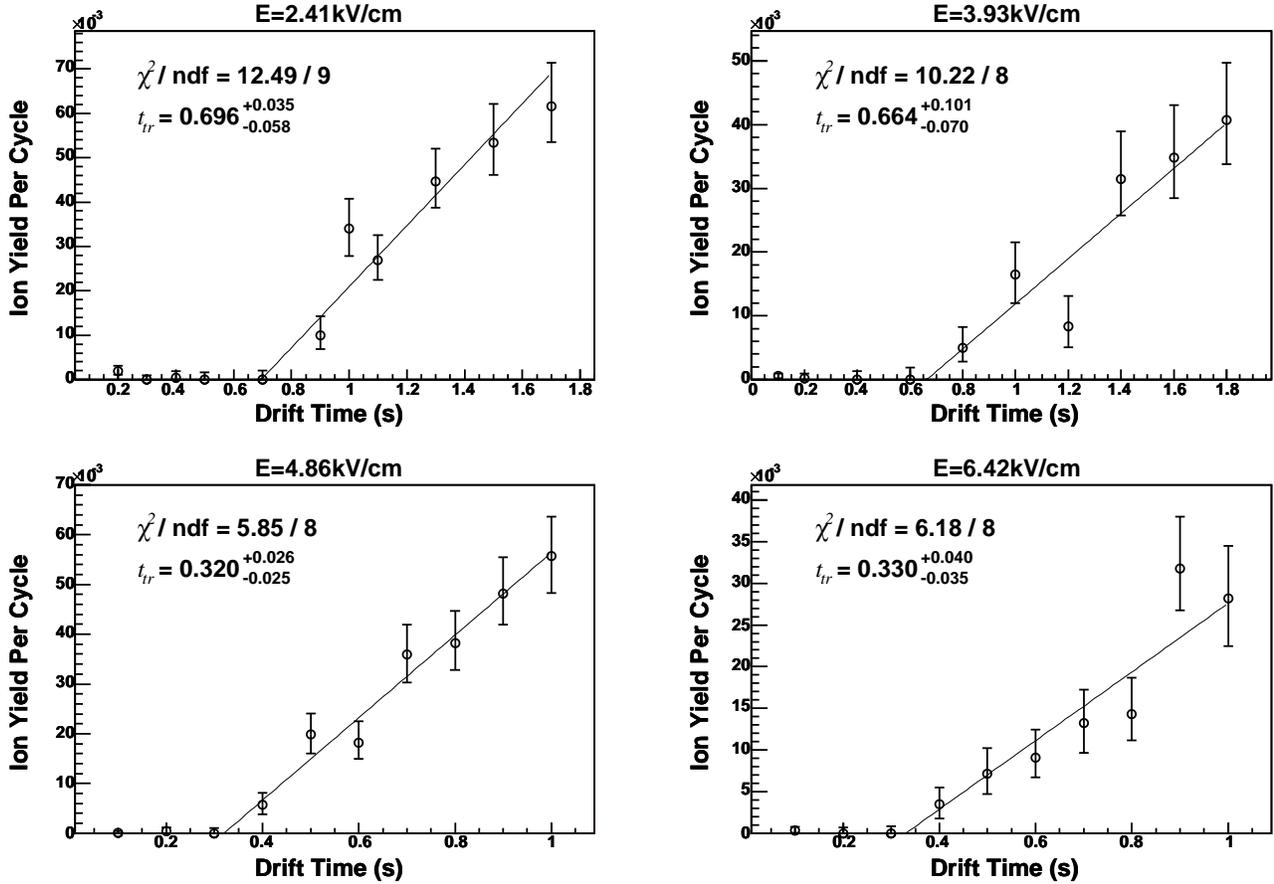}
\caption{
\label{fig:transit} Ion yield $Y$ as a function of drift time $t_{drift}$ for the four electric fields 2.41 kV/cm, 3.93 kV/cm, 
4.86 kV/cm, and 6.42 kV/cm.  
Error bars shown are statistical only.  We take the transit time in each case to be the knee point in the fit.}\end{figure*}

\begin{figure}
  \centerline{
    \includegraphics[width=9cm]{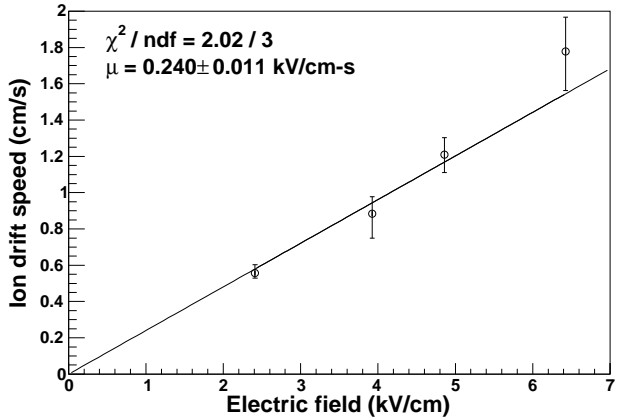}
    }
 \caption{\label{fig:mobilityplot} Global fit to the ion drift velocity as a function of electric field for the entire data set.
     }
\end{figure}

Shown in Fig. \ref{fig:transit} are plots of the ion yield vs $t_{drift}$.  For each data point, the ion yield is 
determined by dividing the number of detected \begin{math}\alpha\end{math} particle counts by the number of 
cycles, giving the ion yield per cycle.  The average background in these units is $\le$0.002,  and the statistical error bounds on
each data point were computed using the Feldman-Cousins prescription.\cite{feld-cous} 
We find that the ion yield is consistent with zero (after background subtraction) when $t_{drift}$ is short, while at longer 
times, the ion yield per cycle increases linearly with $t_{drift}$. The abscissa of the join point between these two behaviors 
is the ion transit time.  We fit the ion yield data to the following two-parameter piecewise 
linear function, \begin{eqnarray*}Y&=&0~~~~~~~~~~~~~~~~~~~~~~t_{drift}<t_{tr}  \\Y&=&m(t_{drift}-t_{tr})~~~~~t_{drift}\geq t_{tr}
\label{exeq}\end{eqnarray*} where $Y$ is the ion yield per cycle, $m$ is the slope (ion yield per cycle per unit $t_{drift}$), 
and $t_{tr}$ is the ion transit time.  The value of $t_{tr}$ yielded by the fit is used to calculate the ion drift velocity.
To perform the fits, and to calculate the statistical error on $t_{tr}$ in each case, we use an algorithm due 
to D.J. Hudson\cite{hud} \begin{table}
\caption{\label{tab:extab} Electric fields (E), voltages ($V_{drift}$), and drift distances ($d$) used in this experiment, showing 
their associated systematic errors.  The measured transit times ($t_{tr}$) 
and velocities ($v$) are the values returned by the fit.  For these last two quantities, only statistical errors are given.}
\begin{ruledtabular}
\begin{tabular}{ccccc}
$d$  &  $V_{drift}$ & E & $t_{tr}$ & $v$ \\
($\times$10 cm) &
($\times$10 kV) &
(kV/cm) &
($\times$10 s) &
($\times$10 cm/s) \\
\hline
\\[-8.5pt]
3.87$\pm$0.08  & 9.32$\pm$0.14  & 2.41$\pm$0.06 & $6.96^{+0.35}_{-0.58}$ & $5.56^{+0.27}_{-0.46}$\\
5.87$\pm$0.08  & 23.10$\pm$0.24 & 3.93$\pm$0.07 & $6.64^{+1.01}_{-0.70}$ & $8.84^{+1.13}_{-0.93}$\\
3.87$\pm$0.08  & 18.88$\pm$0.10 & 4.86$\pm$0.11 & $3.20^{+0.26}_{-0.25}$ & $12.1^{+0.98}_{-0.94}$\\
5.87$\pm$0.08  & 37.71$\pm$0.32 & 6.42$\pm$0.11 & $3.30^{+0.40}_{-0.35}$ & $17.8^{+2.15}_{-1.89}$\\
\end{tabular}
\end{ruledtabular}
\end{table} that is designed specifically to handle the difficulties that arise when performing a piecewise fit.  
One peculiarity of such a fit is that the error bounds are almost always asymmetric. 
\cite{hud} The results of the fits are summarized in Table \ref{tab:extab}.\\ \indent
Fig. \ref{fig:mobilityplot} shows a plot of the computed ion drift velocity as a function of electric field.  The data are 
consistent with the linear relation $v=\mu E$, where $v$ is the ion drift speed, $\mu$ is the mobility, and $E$ is the electric 
field.  From a fit to the data in Fig. \ref{fig:mobilityplot} we find an ion mobility $\mu=$0.240$\pm$0.011 cm$^2$/(kV-s).   This fit 
is obtained by forcing the straight line to go through the origin and it yields $\chi^2$/ndf=2.02/3.A fit 
not constrained to go through the origin results in $\mu=0.268\pm0.031$ cm$^2$/(kV-s), a y-intercept 
of -0.095$\pm$0.088 cm/s, and $\chi^2/$ndf=1.14/3. This is fully consistent with the original result.\\ \indent
The mobility and its statistical error are determined largely by the data taken at the lowest 
electric field.  This point is more precise for two reasons.  First, the ion transit time at this electric field is the longest 
of the four datasets, meaning that the fractional error on the velocity is the smallest. Secondly, in this dataset the ion transit 
time happens to be within $\sim$0.01 seconds of the nearest $t_{drift}$ at which we collected data, and this results in a precise 
measurement.
While, in principle, better accuracy could be obtained by repeating the measurements and always taking a point near this optimal 
time,  the 20.8 day half life of the $^{230}$U source makes it impractical to repeat the entire experiment.  It 
is also true that \nopagebreak the data from which the first point is determined was recorded earlier in the 
lifetime of the $^{230}$U source, resulting in higher statistics.\\ \indent
Systematic uncertainties on the mobility derive from small mechanical vibrations of the probe when the electric field is switched, 
and from the accuracy of the voltage gain in the amplifier.  These contributions are estimated to give a combined error of 
$\pm$ 0.011 cm\begin{math}^2\end{math}/(kV-s).  
Table \ref{tab:extab} shows the absolute errors on the voltages, drift distances and electric fields.
The systematic error on the ion mobility was computed by shifting the electric field and drift distance values by their 
corresponding errors and re-doing the fit.\\ \indent

\section{Discussion}

To help us understand our result, we have considered a transport model due to Hilt, Schmidt and Khrapak\cite{sch} (henceforth referred 
to as HSK).  In this model the microscopic properties of the ion and its detailed 
coupling to the liquid atoms are neglected, and the ionic mobility is determined solely by the equation of state of the liquid in 
question.  The ion is modeled as a point charge whose electric field exerts a pressure on the liquid of sufficient magnitude 
for a microscopic ice layer to form around the ion.  The ion and its attached cluster of Xe atoms then behave as a positively 
charged ball of ice whose radius is determined by the distance at which the
electric field  is no longer of a magnitude sufficient to compress the liquid into the solid state\cite{sch}.  In this theory, 
no distinction is made between the mobilities of ions of different identities, and the drag suffered by the ion and its attached 
cluster is computed from the bulk viscosity of the liquid.\\ \indent
Shown in Table \ref{tab:theory} are some selected experimental mobilities  \begin{table}
\caption{\label{tab:theory} Ion mobilities in LXe, comparing
a model due to Hilt, Schmidt and Khrapak (HSK) \cite{sch} to the 
experimental data.}
\begin{ruledtabular}
\begin{tabular}{ccccc}
\multicolumn{1}{c}{Ion}&\multicolumn{1}{c}{T (K)}&\multicolumn{1}{c}{P (bar)}
&\multicolumn{2}{c}{Mobility  (cm$^{2}$ kV$^{-1}$ s$^{-1}$)}\\ 
& & &
HSK & Expt.\\
\hline
\\[-4.5pt]
$^{208}$Tl$^{+}$&163.0&0.9& 0.31 & 0.133$\pm$0.004\footnotemark[1]\\
$^{226}$Th$^{+}$&163.0&0.9& 0.31 &  0.240$\pm$0.011$\pm$0.011\footnotemark[2]\\
Ba$^{+}$&168.0&1.2& 0.33 &  0.211$^{+0.020}_{-0.012}$\footnotemark[3]\\
\end{tabular}
\end{ruledtabular}
\footnotetext[1]{Ref. \onlinecite{wal}.}
\footnotetext[2]{This work.}
\footnotetext[3]{Ref. \onlinecite{fai}.}
\end{table}for atomic ions in LXe.  From these results it is clear that 
the HSK model does quite well in predicting the overall scale of the observed mobilities.  However, the significant difference 
between the $^{208}$Tl$^{+}$ and $^{226}$Th$^{+}$ numbers suggests that these species differ enough microscopically that the HSK 
model, when applied to them, begins to break down.\\ \indent
It should be pointed out that all of the calculations assume an ionic charge state of +1, which may not be correct.  
This is based on the argument that the energetics of charge exchange favor
such a final state,\cite{wal} but this assertion has not yet been convincingly verified experimentally.  Nonetheless, we find 
that assuming an ionic charge other than +1 does not improve the compatibility between the theory and the data
for any of the cases considered here.\\ \indent
Overall, it can be concluded that HSK provides a reasonable rough approximation of the ion mobility for the systems under 
consideration. 
However, it remains to be seen whether the model produces the correct temperature dependence in all cases.
In addition, it is worth understanding the situations in which the ion identity becomes important because this would offer clues as to 
how HSK could be simply improved.\\ \indent
A theory put forward by Davis, Rice and Meyer (henceforth referred to as DRM) might offer some clues as to how this might be 
accomplished.\cite{dav}  These 
authors have derived an expression for the positive ion mobility in which the internal structure of the ion and its interaction with 
the liquid are taken into account.  In this model the drag force experienced by the ion is determined by the microscopic 
interaction potentials rather than the bulk  properties of the liquid.   It is conceivable, at least in principle, that as the 
ion-liquid interaction increases in strength, bound states between the ion and several liquid atoms can occur, particularly as the 
temperature of the liquid is lowered. Nevertheless, the treatment reported in Ref. \onlinecite{dav} is specific to conditions in which 
this effect is not present, i.e., the ion-liquid interaction is too weak and the liquid temperature is too high.  Indeed, the DRM model
has only been compared with experimental data from systems involving molecular ions in liquids at temperatures well above the 
liquid triple point, namely He$_{2}$$^{+}$ in LHe (Ref. \onlinecite{dav2}), Ar$_{2}$$^{+}$ in LAr (Ref. \onlinecite{dav}), and 
Kr$_{2}$$^{+}$ in LKr (Ref. \onlinecite{pal}).  
While these systems are satisfactorily described by the model, it is clear that some modification will be necessary in order to 
treat the case of atomic ion mobilities close to the triple point.  This is because the atomic ions under consideration have 
interaction strengths that are tens of times stronger\cite{kir,mas} than those of the molecular ions previously considered.\\ \indent
To investigate these details experimentally, one could perform an accurate measurement of the temperature dependence of the ion 
mobility in several different ion-liquid
systems. This would require multiple, well-understood and well-controlled sources of ions of known identity.  A complementary approach 
would be to 
use the $^{226}$Th ion source in several different liquefied noble gases at various temperatures.\\ \indent

\section{Conclusion}
We have measured the $^{226}$Th ion mobility in LXe at 163.0K and standard pressure and have found it to have 
a value of 0.240\begin{math}\pm\end{math}0.011 (stat) \begin{math}\pm\end{math}0.011 (syst) cm\begin{math}^{2}\end{math}/(kV-s).
The HSK model gives the correct order of magnitude when used to calculate ion mobilities, however it cannot account for the 
significant difference between our result and the one measured for $^{208}$Tl$^+$ by other workers.  A more detailed study of this 
effect may help to suggest how the HSK model can be modified to account for different ion identities.  A possible route might be to 
generalize the DRM treatment to the case of atomic ions that form bound states with liquid atoms.\\

\section{Acknowledgements}
The authors wish to thank K. Roberts of Lawrence Livermore National Laboratory for his assistance in manufacturing the 
$^{230}$U source used in the experiment.  This work is supported by the U.S. Department of Energy.

%
\end{document}